\documentclass[usenatbib,useAMS]{mn2e}
\usepackage{graphicx}
\usepackage{epstopdf}
\usepackage{ctable}
\usepackage{url}
\usepackage{times}

\newcommand{\lir}{L_{\rm IR}}

\newcommand{\msun}{~\mathrm{M_{\odot}}}
\newcommand{\msunperyr}{~\mathrm{M_{\odot} {\rm ~yr}^{-1}}}

\newcommand{\mstar}{M_{\star}}
\newcommand{\mdust}{M_\mathrm{d}}

\newcommand\dllg{\delta_{\mathrm{LBG}}}
\newcommand\dsmg{\delta_{\mathrm{SMG}}}
\newcommand\dm{\delta_{\mathrm{mass}}}
\newcommand\dgal{\delta_{\mathrm{galaxy}}}

\newcommand{\acknowledgments}{\begin{small}\section*{Acknowledgements}\end{small}}

\newcommand\fref[1]{\hyperref[#1]{Fig.~\ref*{#1}}}
\newcommand\tref[1]{\hyperref[#1]{Table~\ref*{#1}}}

\title[SMGs as tracers of large-scale structure]{The bias of the submillimetre galaxy population:~SMGs are poor tracers of the most massive structures in the $z\sim2$ Universe}

\author[Miller, Hayward, Chapman \& Behroozi]{
\parbox[t]{\textwidth}{
Tim B. Miller$^{1}$\thanks{E-mail: tim.blake.miller@gmail.com},
Christopher C. Hayward$^{2,3}$\thanks{Moore Prize Postdoctoral Scholar in Theoretical Astrophysics},
Scott C. Chapman$^{1}$ and Peter S. Behroozi$^{4}$\thanks{Giacconi Fellow}}
\vspace*{6pt} \\
$^1$Department of Physics and Atmospheric Science, Dalhousie University, 6310 Coburg Road, Halifax, NS B3H 4R2, Canada\\
$^2$TAPIR 350-17, California Institute of Technology, 1200 E. California Boulevard, Pasadena, CA 91125, USA\\
$^3$Harvard--Smithsonian Center for Astrophysics, 60 Garden Street, Cambridge, MA 02138, USA\\
$^4$Space Telescope Science Institute, 3700 San Martin Drive, Baltimore, MD 21218, USA
}

\usepackage[breaklinks,colorlinks,allcolors=blue]{hyperref}
\usepackage[all]{hypcap}
\begin{document}

\date{Accepted for publication in MNRAS}

\pagerange{\pageref{firstpage}--\pageref{lastpage}} \pubyear{2015}

\maketitle

\label{firstpage}

\begin{abstract}
It is often claimed that overdensities of (or even individual bright) submillimetre-selected galaxies (SMGs) trace the assembly of the most-massive
dark matter structures in the Universe.
We test this claim by performing a counts-in-cells analysis of mock SMG catalogues derived from the \emph{Bolshoi} cosmological simulation to
investigate how well SMG associations trace the underlying dark matter structure. We find that SMGs exhibit a relatively complex bias:
some regions of high SMG overdensity are underdense in terms of dark matter mass, and some regions of high dark matter overdensity contain no SMGs. 
Because of their rarity, Poisson noise causes scatter in the SMG overdensity at fixed dark matter overdensity.
Consequently, rich associations of less-luminous, more-abundant galaxies (i.e. Lyman-break galaxy analogues) trace the highest dark matter
overdensities much better than SMGs. Even on average, SMG associations are relatively poor tracers of the most significant
dark matter overdensities because of `downsizing': at $z \la 2.5$, the most-massive galaxies that reside in the highest dark matter
overdensities have already had their star formation quenched and are thus no longer SMGs. At a given redshift,
of the 10 per cent most-massive overdensities, only $\sim 25$ per cent contain at least one SMG, and less than a few per cent contain more than one SMG.
\end{abstract}


\section{Introduction} \label{S:intro}

Submillimetre-selected galaxies (SMGs; see \citealt{Casey2014} for a recent review), with typical infrared (IR) luminosities of $L_{\rm IR} \ga 5
\times 10^{12}$\,L$_{\odot}$, represent the rarest and most extreme examples of star forming galaxies.
The $\lir$ of an SMG implies an immense star formation rate, typically $\rm SFR\sim500-1000$\,M$_{\odot}$\,yr$^{-1}$,
assuming that there is not a significant contribution to $\lir$ from deeply obscured AGN.
SMGs allow us to probe the mechanisms behind most intense star formation events in the Universe and can elucidate the highest SFRs
sustainable in a galaxy. Because of their extreme nature, SMGs provide laboratories to test the limits of hydrodynamical simulations of galaxies
\citep[e.g.][]{Narayanan:2010smg,Hayward:2011smg_selection,Hayward:2013limits}.

Massive starburst galaxies appear to grow in the most massive halos \citep{Hickox:2012}, thus making them potential tracers for the highest-redshift
proto-clusters \citep[e.g.][]{Capak:2011,Walter:2012}. Thus, observations of bright SMGs should probe their environment and trace significant
overdensities that can be interpreted in the context of large-scale structure simulations. Therefore, SMGs set critical constraints on cosmological models.

Interest in SMG associations has grown in recent years as increasing numbers of SMG associations have been detected
\citep[e.g.][]{Blain:2004,Chapman:2005,Chapman:2009,Geach:2005,Daddi:2009smgb,
Dannerbauer2014,MacKenzie2014,Smail2014}.
Furthermore, \citet{Clements2014} have demonstrated that some \emph{Planck} sources trace overdensities of dusty star-forming galaxies,
and they suggested that such observations can be used to investigate the epoch of galaxy cluster formation. However, this claim relies on the assumption
that overdensities of dusty star-forming galaxies correspond to galaxy clusters in the process of formation.

There is some observational evidence that calls this claim into question: in their study of the GOODS-N field, \citet{Chapman:2009}
found an association of 8 SMGs at $z \approx 1.99$.
The associated structure was only a typical overdense region, as indicated by the
well-sampled optical spectroscopy in this region, that would not form a virialized cluster by $z = 0$.
Moreover, \citet{Blain:2004} found that the clustering length of SMGs is consistent with that of evolved `extremely red objects' (EROs) at $z \sim 1$
and $z = 0$ clusters, which would suggest that the descendants of SMGs would tend to be found in rich cluster environments; however, this
interpretation implies a comoving space density of clusters that is at least an order of magnitude greater than that observed.
These results suggest that perhaps associations of SMGs trace particularly active phases in relatively modest-mass overdensities
rather than the highest overdensities and thus have a relatively complex clustering bias.

To test this possibility, we have performed a counts-in-cells analysis on the \citet[][hereafter H13]{HB2013} simulated SMG catalogues to
investigate the relationships of SMGs and more modestly star-forming galaxies to the underlying dark matter structure. We first investigate
the clustering biases of SMGs and Lyman-break-galaxy (LBG) analogues. We then study the properties of individual associations of SMGs
and LBG analogues.

\section{Methods} \label{S:methods}

To analyze the bias in the SMG population, we use the mock SMG catalogues of H13, which were generated by
assigning galaxy properties to dark matter haloes from a cosmological collisionless dark matter simulation using subhalo abundance matching
and then assigning submm flux densities using a fitting function derived from the results of performing radiative transfer on idealized
hydrodynamical simulations. We will summarize the H13 methodology here, and we refer the reader to H13 for full details.

Using halo catalogues from the \textit{Bolshoi} simulation (\citealt*{Klypin:2011}; \citealt{Behroozi:2013rockstar,Behroozi:2013}), we constructed
mock lightcones by starting at eight random locations within the simulation and selecting haloes along a randomly oriented sightline with an
84' x 84' (1.96 deg$^2$) field of view from $z=0.5$ to $z=8$. We calculated cosmological redshifts, including the effects of halo peculiar velocities.
We then assigned stellar masses ($\mstar$) and SFRs using the redshift-dependent stellar mass--halo mass and SFR--halo mass relations of
\citet*{Behroozi:2013SFH}, which include scatter at fixed halo mass and redshift. We included a simple model for satellite quenching: satellite SFRs
were reduced by a factor equal to their current subhalo mass divided by the peak mass in their subhalo's mass accretion history. We assigned
dust masses ($\mdust$) to the haloes using the empirically based method of \citet{Hayward:2013number_counts}. Finally, we assigned 850-$\mu$m flux densities
($S_{850}$) using the following fitting function, which was derived based on the results of performing dust radiative transfer on hydrodynamical
simulations of idealized disc galaxies and mergers \citep{Hayward:2011smg_selection,Hayward:2013number_counts}:
\begin{equation}
S_{850} = 0.81 {\rm ~mJy} \left(\frac{\mathrm{SFR}}{100 ~\msunperyr}\right)^{0.43} \left(\frac{\mdust}{10^8 \msun}\right)^{0.54},
\end{equation}
where we incorporated the scatter in the relation of 0.13 dex \citep{Hayward:2011smg_selection} when assigning $S_{850}$.
Note that because $S_{850}$ scales sublinearly with both SFR and $\mdust$, the predicted $S_{850}$ values are relatively insensitive to the model details.
Furthermore, the $S_{850}$--$\mstar$ relation predicted in this manner agrees well with that observed \citep{Davies:2013}.

\begin{figure}
\centering
\includegraphics[width=\columnwidth]{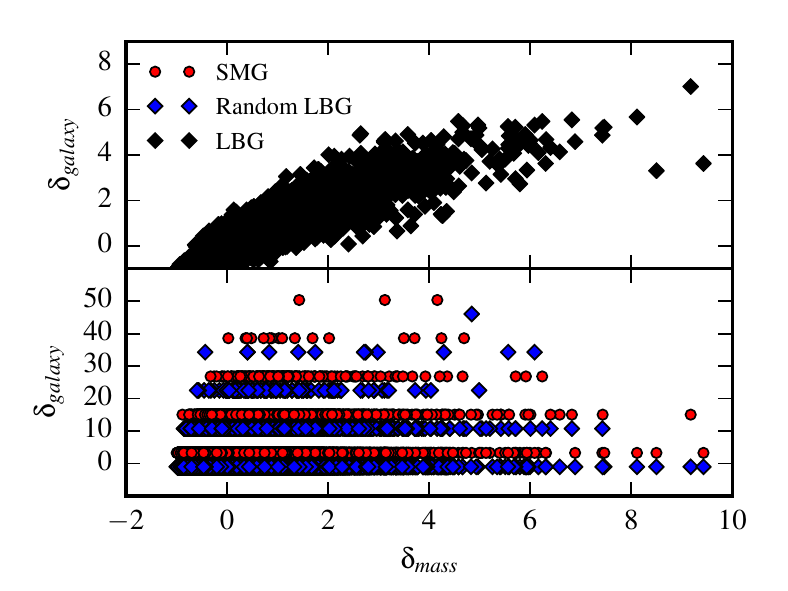}
\caption{\emph{Top:} Overdensity of LBGs (0.1 mJy $< S_{850} <$ 1 mJy) vs. overdensity of dark matter, $\dm$, for the random cells. 
There is a clear correlation between $\dgal$ and $\dm$, which indicates that the clustering of the LBGs traces the clustering of the dark matter well.
\emph{Bottom:} Similar to the top panel, but for SMGs ($S_{850} > 3$ mJy; red circles) and a random subset of LBGs selected to have number
density equal to that of the SMGs (blue diamonds). (The SMG points have been shifted upwards by 5 for clarity.) 
The SMGs and random subset of LBGs exhibit a large scatter in $\dgal$ at a given $\dm$. This result indicates that because of the
rarity of SMGs, Poisson noise causes SMG overdensities to be poor tracers of dark matter overdensities.}
\label{fig:od}
\end{figure}

Throughout this work, we refer to mock galaxies with $S_{850} > 3$ mJy as SMGs (the median SFR for sources with $S_{850} \sim 3$ mJy
is $\sim 140 \msunperyr$) and those with $0.1 < S_{850} < 1$ mJy as LBGs (this range corresponds to median SFR values of
$\sim 10-50 \msunperyr$). We study the bias of SMGs and LBGs and identify SMG and LBG associations (or redshift spikes;
e.g. \citealt{Chapman:2005,Chapman:2009})
using a simple counts-in-cells analysis \citep[e.g.][]{Adelberger:1998}. Specifically, we divide each of the 8 mock catalogues into cells with
angular dimensions 10 arcmin $\times$ 10 arcmin and depth $dz = 0.05$; the results are similar if we use cells with side lengths equal to twice these
values. We use a subset of 10,000 of these cells for calculating the clustering bias and for making comparisons to the properties of SMG and LBG
associations; we refer to these cells as `random cells'. To identify associations, we start with the same cells. However, to ensure that we do not
divide potential associations by using a fixed grid, we shift the cells by 1-arcmin intervals 10 times and define an SMG (LBG) association as the
galaxies contained in the cell that contains the maximum number of SMGs (LBGs). We ensure that we do not count a single association multiple times.
We calculate total dark matter masses for each cell by summing the dark matter masses of all haloes of mass $> 10^{10} \msun$ (because the
\emph{Bolshoi} simulation is incomplete below this halo mass) contained in the cell.

\begin{figure*}
\centering
\includegraphics[width=0.9\columnwidth]{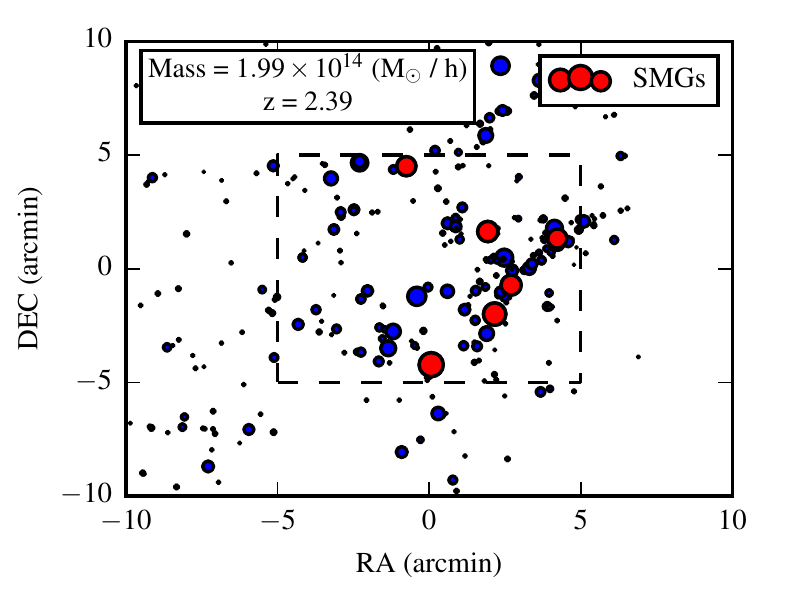}
\includegraphics[width=0.9\columnwidth]{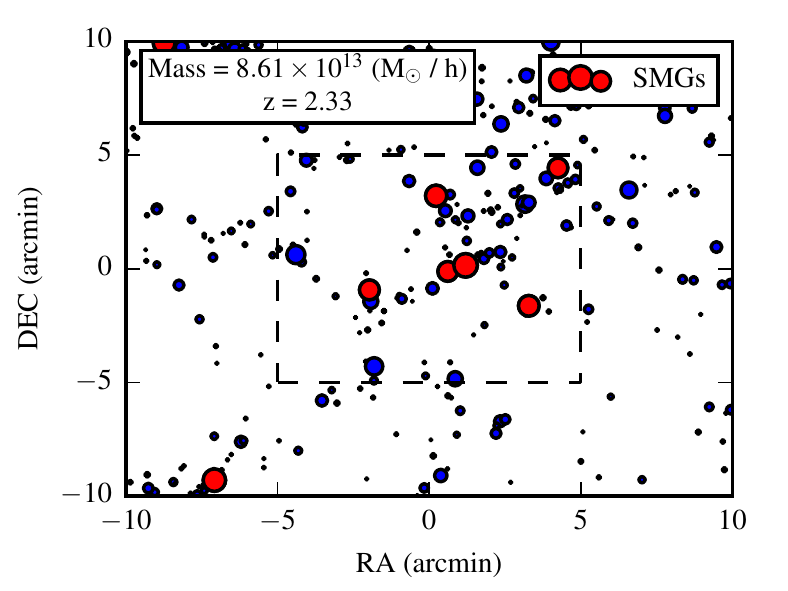}
\caption{Spatial distributions of galaxies near the 2 richest SMG associations (which each contain 6 SMGs; the cells are marked with dashed lines).
The SMGs (LBGs) are denoted with red (blue) points, the sizes of which are proportional to S$_{850}$. The spatial distributions of the galaxies reflect
the filamentary structure of the dark matter distribution.}
\label{fig:location}
\end{figure*}

As discussed in detail in H13, the H13 model does not include the effect of starbursts (i.e. the extended tail to high SFR at a given
stellar mass and redshift). Because one would expect that interaction-induced starbursts would most affect the SFRs of SMGs
in highly overdense regions and thus potentially alter our results, we have extended the H13 model by including a model for
interaction-induced starbursts. For each galaxy, we check whether it has a neighboring galaxy with stellar mass within a factor of 3 of
its own (such that the pair would constitute a `major' merger) that is located within a physical distance of $d_{\mathrm{weak}}$. If so, the
SFRs of both galaxies are boosted by a factor of $b_{\mathrm{weak}}$. If the separation is less than $d_{\mathrm{strong}} < d_{\mathrm{weak}}$, we
instead boost the SFRs by a larger factor, $b_{\mathrm{strong}} > b_{\mathrm{weak}}$.
We experimented with different reasonable parameter values, as judged based on the results of idealized hydrodynamical simulations of mergers
\citep[e.g.][]{Cox:2008,Torrey:2012,Hayward2014arepo}, and found that the results were qualitatively unaffected even for the extreme
scenario of $d_{\mathrm{weak}} = 15$ kpc, $b_{\mathrm{weak}} = 10$, $d_{\mathrm{strong}} = 5$ kpc, and $d_{\mathrm{strong}} = 100$.
However, it is important to note that the catalogues are incomplete for mergers with small separations \citep{Behroozi:2013rockstar},
which would result in an underestimate of the number of interacting galaxies. Nevertheless, this incompleteness likely does not
affect our results because although interactions could boost the submm fluxes of some galaxies and increase the clustering signal on
$\la 10$ kpc scales, the clustering on larger scales should be unaffected.
In all figures, we show the results for the original H13 model, but the corresponding plots for the boosted models are similar.

Note that unlike H13, we have not incorporated the effects of blending of multiple galaxies into a single submm source in this work,
although theoretical arguments (\citealt{Hayward:2011smg_selection,Hayward:2012smg_bimodality,
Hayward:2013number_counts}; H13) and observations \citep[e.g.][]{Karim2013,Hodge2013} suggest that blending
significantly affects the single-dish-detected SMG population.
The reason is that the sizes of the associations are much greater than the beam sizes of single-dish submm telescopes (see below).
Thus, although the detailed results
could be affected by blending, our conclusions would be unchanged if blending were incorporated. Furthermore, we wish to analyze
how well individual submm-bright
galaxies, which would be resolved by e.g. the Atacama Large Millimeter Array, rather than blended submm sources (which depend
on the beam size of the instrument used to detect them and are thus a less general population than resolved sources) trace
dark matter structures.

\section{Results} \label{S:results}

For each cell, we calculate the number overdensity of SMGs ($\dsmg$) and LBGs ($\dllg$) using the following equation:
\begin{equation}
\dgal = \frac{N_{\mathrm{galaxy}} - <N_{\mathrm{galaxy}}>}{<N_{\mathrm{galaxy}}>},
\end{equation}
where $N_{\mathrm{galaxy}}$ is the number of galaxies in a cell and $<N_{\mathrm{galaxy}}>$ is the mean number of galaxies
per cell. We also calculate the dark matter mass overdensity of each cell,
\begin{equation}
\dm = \frac{M_{\mathrm{DM}} - <M_{\mathrm{DM}}>}{<M_{\mathrm{DM}}>},
\end{equation}
where $M_{\mathrm{DM}}$ is the mass of dark matter in a cell and $<M_{\mathrm{DM}}>$ is the mean dark matter mass per cell.

\fref{fig:od} shows the overdensity $\dgal$ of SMGs ($S_{850} > 3$ mJy; red circles), LBGs (0.1 mJy $< S_{850} <$ 1 mJy; black diamonds) and
a random subset of LBGs selected to have number density equal to that of the SMGs (blue diamonds) vs. overdensity of dark matter, $\dm$, for the random cells.
For the total LBG population, $\dgal$ and $\dm$ are tightly correlated, which indicates that LBG overdensities are good tracers of dark matter overdensities.
The slope of the best-fitting linear relation (i.e. the bias, $b \equiv \dgal/\dm$) is $0.98 \pm 0.01$, and the mean squared error (MSE) is 0.1.
For the SMGs and random subset of LBGs, there is a correlation between $\dgal$ and $\dm$, but it exhibits significant scatter. The bias values are
$1.3 \pm 0.1$ and $1.0 \pm 0.1$ for the SMGs and LBGs, respectively, and the MSE values are 17 and 16. The fact that the SMGs
and random LBGs exhibit similar scatter indicates that Poisson noise due to the rarity of SMGs is the reason for the complicated relationship between
$\dgal$ and $\dm$ for this population. Thus, although SMGs are slightly more biased than LBGs, \emph{SMG overdensities are poor tracers of the underlying
dark matter overdensities.} This effect may explain the results of \citet{Blain:2004} discussed above.

\begin{figure*}
\centering
\includegraphics[width=0.9\columnwidth]{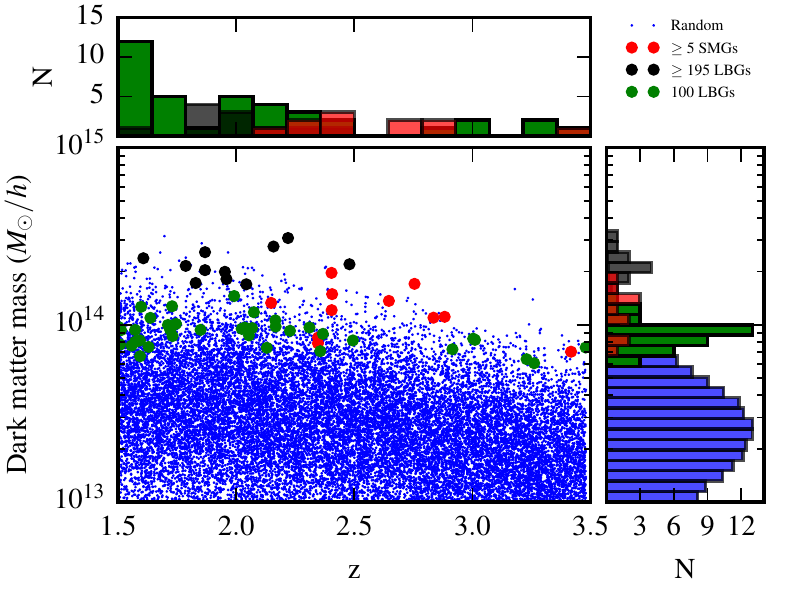}
\includegraphics[width=0.9\columnwidth]{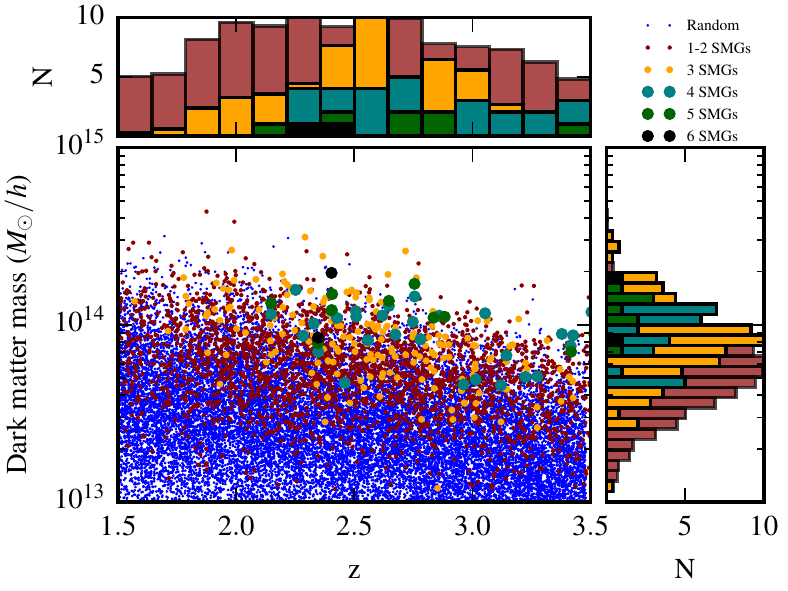}
\caption{\emph{Left:} total dark matter mass in a cell versus redshift of the cell for cells that contain $\ge 5$ SMGs (red circles), cells with $\ge 195$ LBGs (this
number was  selected to yield the 10 richest LBG associations; black circles), cells that contain exactly 100 LBGs (green circles) and randomly selected cells (blue
points). The redshift and dark matter mass distributions are shown next to the respective axes.
Compared with the richest LBG associations, the SMG associations trace less-massive, higher-redshift structures. Associations of 100 LBGs trace lower-mass
dark matter substructures that span the full redshift range considered. \emph{Right:} similar to the left panel, but with cells classified according to the number of
SMGs that they contain. Cells with lower numbers of SMGs tend to include less dark matter. Notably, some of the
most overdense regions contain no SMGs.}
\label{fig:dm_vs_z_comp}
\end{figure*}

\ctable[
	caption = 		{Demographics of SMG associations\label{tab:associations}},
				center,
				notespar
]{lccc}{
	\tnote[a]{Number of SMGs in a cell.}
	\tnote[b]{Percentage of SMGs in such associations.}
	\tnote[c]{Median mass of such associations at $2 < z < 3$.}
	\tnote[d]{Median pairwise separations of SMGs in such associations.}
	}{
																									\FL
$N_{\rm SMG}$\tmark[a]			&	Percentage\tmark[b]		& Mass\tmark[c]		&	Separation\tmark[d]			\NN
							&						& ($10^{13} \msun$)		&	(Mpc)					\ML
1							&	64					& $5.9 \pm 3.0$		&	--						\NN
2							&	23					& $7.7 \pm 3.6$		&	7.6						\NN
3							&	10					& $9.1 \pm 4.7$		&	6.1						\NN
4							&	2					& $10.3 \pm 2.8$		&	5.2						\NN
5							&	0.8					& $12.6 \pm 2.6$		&	4.3						\NN
6							&	0.2					& $14.1 \pm 5.6$		&	4.6						\LL
}

We now investigate the properties of individual SMG and LBG redshift associations in detail.
\tref{tab:associations} presents the fraction of SMGs in associations and the median separation of the SMGs in the different types of
associations. Typical SMGs are not in associations, but a substantial minority (36 per cent) are. Only a few per cent of SMGs are located
in rich associations of four or more SMGs. The richer associations exhibit lower median separations, which suggests that these
associations correspond to higher overdensities (which will be confirmed below).
\fref{fig:location} shows the spatial distributions of the SMGs and LBGs in two mock SMG associations. The associations each contain 6 SMGs.
The spatial distributions of the galaxies exhibit clear filamentary structures, which reflect the underlying structure of the
`cosmic web'. Note that incorporating blending (using a typical beam of $\la 30$ arcsec) would tend to increase the
submm fluxes of the LBGs and blend the SMGs with LBGs, but it would not blend any of the bright SMGs. Thus, a blended version
would potentially contain additional bright SMGs and therefore be comparable to observed SMG associations \citep[e.g.][]{Chapman:2009,Dannerbauer2014}.

Our goal is to understand how well SMG associations trace the highest dark matter overdensities. To do so, it is instructive to compare
the total dark matter mass in individual cells that are identified as SMG or LBG associations with the values for randomly selected cells. If SMG associations
trace the most significant overdensities, these cells should contain more dark matter mass than other cells at a given redshift.
The left panel of \fref{fig:dm_vs_z_comp} shows the total dark matter in a given cell versus the redshift of the cell for cells that contain $\ge 5$
SMGs (red circles), the 10 richest LBG associations ($\ge 195$ LBGs in a cell; black circles), cells that contain exactly 100 LBGs (green circles),
and a subset of randomly selected cells (blue points). The redshift and dark matter mass distributions are shown next to the respective axes.

It is immediately clear that the \emph{SMG associations do not trace the most significant dark matter overdensities}, although they do trace relatively
high overdensities. Compared with the richest LBG associations, which should be considered analogous to observed associations (i.e. redshift spikes) of LBGs,
the SMG associations tend to have lower dark matter masses (a median of $1.2 \times 10^{14} \msun$ for the SMG associations compared with $2.2 \times 10^{14}
\msun$ for the $\ge 195-$LBG associations) and are located at higher redshifts (the median values for the SMG and LBG associations are 2.4 and 2.0, respectively).
Furthermore, there are many randomly selected cells that have dark matter masses that are comparable to or even greater than the values for the SMG associations,
whereas the richest LBG associations more faithfully trace the cells with the largest dark matter masses.

For comparison, we show more-modest LBG associations that contain exactly 100 LBGs (green circles). As expected, these LBG associations trace less massive
substructures than the richest LBG associations. The median dark matter mass of the 100-LBG associations is similar to that of the SMG associations, $9.1 \times 10^{13}
\msun$, but the 100-LBG associations span a broader range of redshifts.

The right panel of \fref{fig:dm_vs_z_comp} shows the total
dark matter mass in a cell versus redshift of the cell for cells that contain one or more SMGs (coloured according the number of SMGs).
This figure demonstrates multiple interesting results: first, \emph{many of the highest dark matter overdensities at a given redshift contain no SMGs}
(the blue points with dark matter mass $\ga 2 \times 10^{14} \msun$). At a given redshift, of the 10 per cent most-massive overdensities, only $\sim 25$
per cent contain at least one SMG, and less than a few per cent contain more than one SMG.
Consequently, finding dark matter overdensities using SMGs as signposts will cause one to miss many of the highest overdensities.
Second, cells with lower numbers of SMGs tend to have less dark matter. Finally, the minimum mass necessary for a cell to
host an SMG is $\sim 10^{13} \msun$, which is consistent with the results of inferences from the clustering of real SMGs \citep{Hickox:2012}.

\section{Summary and Discussion}

We have used mock SMG catalogues to demonstrate that SMG associations are poor tracers of the highest overdensities of
dark matter. At higher redshifts ($z \ga 2.5$), the richest SMG associations trace some of the highest overdensities because the most-massive galaxies in those regions
are still forming stars rapidly. However, such associations are rare, and the majority of the highest overdensities do not host a single SMG,
let alone an SMG association. Consequently,
SMG associations are highly incomplete tracers of the highest overdensities even at $z \ga 2.5$.
The situation is worse at $z \la 2.5$: many of the most-massive galaxies, which reside in the highest dark matter overdensities, have already had their star
formation quenched. (Independently of redshift, the halos with the highest
ratio of SFR to halo mass are those with halo masses of $\sim 10^{12} \msun$ at that redshift; e.g. \citealt{Behroozi:2013SFH,Moster:2013,
Sparre2015}.) Consequently, the $z \la 2.5$ dark matter overdensities are less likely to contain SMGs.

In our model, galaxy SFRs are assigned using a redshift-dependent SFR--halo mass relation and a model for satellite quenching. The parameters of the model
are constrained by fitting to a wide range of observations \citep{Behroozi:2013SFH}. Consequently, the fact that some fraction of massive galaxies are
quenched even at $z \sim 2$ is not a prediction. The utility of our model is that
it can be used to determine the consequences of quenching/downsizing for the clustering of the SMG population. Furthermore, because we determine
the submm flux densities of our galaxies self-consistently using a fitting function derived from radiative transfer calculations, there is not a monotonic mapping
between SFR and submm flux density (a galaxy with a relatively modest SFR can still be submm-bright if it has sufficiently high dust mass). Thus, the
results are specific to the SMG population rather than just the most rapidly star-forming galaxies (cf. \citealt{Dave:2010}). Finally, our model explicitly
accounts for the stochasticity that is inherent in the SMG selection because bright SMGs are an extreme population; thus, one may select a galaxy as an
SMG because it is in a short-lived phase of elevated SFR (perhaps due to an interaction) or because it has an especially high submm flux density for its
SFR and dust mass (because of e.g. an especially extended geometry). GN20 could be a real-Universe example of the latter.
Consequently, Poisson noise contributes to the scatter in the value of $\dsmg$ at a given
$\dm$ and causes some of the most significant overdensities to contain few or no bright SMGs. Moreover, our model suggests that some the brightest
SMGs in the Universe may lie in relatively isolated massive dark matter halos, consistent with observational findings \citep{Chapman2015}.

A few other theoretical works have investigated the clustering of the SMG population. \citet{Dave:2010} studied the properties of the most rapidly star-forming
galaxies, which they considered SMG analogues, in a cosmological hydrodynamical simulation. Because of the tight, monotonic SFR--stellar mass relation for
star-forming galaxies in their simulation, they effectively selected the most-massive star-forming galaxies in their simulation. Consequently, they found
that their simulated SMGs were highly clustered and biased, with a correlation length $r_0 \approx 10 h^{-1}$ comoving Mpc and bias of $\sim 6$.

\citet{Cowley2015b} analyzed the clustering of SMGs in the current Durham semi-analytical model (\citealt{Cowley2015a}; Lacey et al., in preparation).
In this model, bright ($S_{850} > 4$ mJy) SMGs at $z \sim 2.5$ exhibit a correlation length of $r_0 = 5.5 h^{-1}$ Mpc and a bias of $\sim 2.5$. Both
the correlation length and bias are almost independent of $S_{850}$ (at least for $S_{850} > 0.25$ mJy), which is qualitatively consistent with our results
(i.e. galaxies in high overdensities are not necessarily submm-bright). Interestingly, similar results were obtained for a previous version of the model
in which the physical nature of the SMGs was qualitatively very different \citep{Almeida:2011}.\footnote{In the current model, SMGs are predominantly
starbursts driven by disk instabilities, and a mildly top-heavy initial mass function is used in starbursts. In the previous model, SMGs were predominantly
very gas-rich galaxies undergoing starbursts driven by minor mergers, and a flat IMF was assumed for starbursts.} However, the current model better
matches the angular power spectrum of cosmic infrared background (CIB) anisotropies \citet{Cowley2015b} than did the previous model,
which suggests (perhaps unsurprisingly) that the CIB may provide more insight into the nature of SMGs than clustering does.

\citet{Granato2014} performed dust radiative transfer on simulated (proto-)clusters at $z \ga 1$ to determine whether they could reproduce the properties
of the \emph{Planck} sources of \citet{Clements2014}. They found that their simulated (proto-)clusters had lower total SFRs than inferred from observations.
This tension suggests that associations of dusty galaxies should be investigated further both in terms of theory and observation.

It is worthwhile to clarify why we have claimed that SMGs are `poor tracers' of high overdensities. In our model, the $\ge 5$-SMG associations each lie in
one of the most-massive cells at their respective redshifts. In this sense, SMG associations do trace high overdensities. However, such associations are uncommon
(there are 11 in our $\sim16-$deg$^2$ mock catalogue), which is consistent with the fact that only a few overdensities of $\ge5$ bright SMGs have been
reported in the literature \citep{Tamura2009,Chapman:2009,Dannerbauer:2009,Ma2015}. Moreover, at $z \la 2.5$, almost none of the highest overdensities
contain $>3$ SMGs despite a large fraction of the SMG population being located at $z \la 2.5$. Thus, searching for high overdensities using SMG associations
as beacons would result in a very incomplete sample at best.

Overall, our results urge caution when interpreting SMG associations in the context of large-scale structure. Because of their rarity, Poisson
noise causes significant scatter in the SMG overdensity at fixed dark matter overdensity (i.e. SMGs stochastically sample the highest overdensities).
Consequently, although
the highest-redshift SMG associations trace some of the highest dark matter overdensities at those redshifts, most of the highest
overdensities do not host SMG associations. At lower redshifts ($z \la 2.5$), the situation is worse: the highest overdensities tend to contain only a few SMGs at most,
and the majority do not contain a single SMG. Thus, if one wishes to identify protoclusters, the complicated bias of SMGs makes them less-than-ideal beacons.

\acknowledgments

We thank Neal Katz for useful discussion and Phil Hopkins for comments on the manuscript. We thank the anonymous referee for a constructive report that
helped improve the manuscript. CCH is grateful to the Gordon and Betty Moore
Foundation for financial support and acknowledges the hospitality of the Aspen Center for Physics, which is supported by the National Science
Foundation Grant No. PHY-1066293. PSB was supported by a Giacconi Fellowship provided through the Space Telescope Science Institute,
which is operated by the Association of Universities for Research in Astronomy under NASA contract NAS5-26555.
\\

\footnotesize{
\bibliography{std_citations,smg}
}

\label{lastpage}

\end{document}